\RequirePackage{fix-cm}
\documentclass{svjour3}                     
\smartqed  
\usepackage{graphicx}
 \usepackage{mathptmx}      
 \journalname{Meccanica}

\usepackage{graphicx}
\usepackage{xcolor}
\usepackage{amsmath}
\usepackage{amssymb}
\usepackage{bm}
\usepackage{ar}
\usepackage{changes}

\newcommand{\etal}{et al.\ }

\newcommand{\eg}{e.g.,\ }

\definecolor{cinnamon}{rgb}{0.82, 0.41, 0.12}

\begin{document}

\title{Transport and evaporation of virus-containing droplets exhaled by men and women {in typical cough events}}


\author{Stefano Olivieri         \and
        Mattia Cavaiola \and
        Andrea Mazzino \and
        Marco E. Rosti
}


\institute{Stefano Olivieri \and Marco E. Rosti \at
              Complex Fluids and Flows Unit, Okinawa Institute of Science and Technology Graduate University, 1919-1 Tancha, Onna-son, Okinawa 904-0495, Japan \\
              \email{marco.rosti@oist.jp}            
           \and
           Mattia Cavaiola \and Andrea Mazzino \at
              INFN and Department of Civil, Chemical and Environmental Engineering (DICCA), University of Genova, Via Montallegro 1, 16145, Genova, Italy
}

\date{Received: date / Accepted: date}

\maketitle

\begin{abstract}
{
The spreading of the virus-containing droplets exhaled during respiratory events, e.g., cough, is an issue of paramount importance for the prevention of many infections such as COVID-19. According to the scientific literature, remarkable differences can be ascribed to several parameters that govern such complex and multiphysical problem. Among these, a particular influence appears associated with the different airflows typical of male and female subjects. Focusing on a typical cough event, we investigate this aspect by means of highly-resolved direct numerical simulations of the turbulent airflow in combination with a comprehensive Lagrangian particle tracking model for the droplet motion and evaporation. We observe and quantify major differences between the case of male and female subjects, both in terms of the droplet final reach and evaporation time. Our results can be associated with the different characteristics in the released airflow and thus confirm the influence of the subject gender (or other physical properties providing different exhalation profiles) on both short-range and long-range airborne transmission. }
\end{abstract}

\section{Introduction} \label{sec:introduction}

In light of the on-going COVID-19 pandemic, the understanding of how \sloppy{virus-containing droplets exhaled during respiratory events (e.g., when coughing, sneezing or talking)} travel in the surrounding environment has gained considerable importance~\cite{bahl2020airborne,lincei2020review,mittal2020flow}. An intense debate  has raised within the scientific community on the possibility that airborne transmission could play a relevant role in the spreading of SARS-CoV-2 and similar infections~\cite{morawska2020time,lewis2020coronavirus,asadi2020coronavirus}, resulting in a renovated research effort in the analysis of this complex and cross-disciplinary problem which includes experimental tests~\cite{lee2019quantity,bahl2020experimental} as well as investigations of computational nature~\cite{balachandar2020host,chong2021extended,rosti2020turbulence,rosti2020urgent,wang2021short}. 
 
In the latter framework, in order to simulate expiratory events in an accurate way, firstly we have to select proper and representative initial conditions. This means to characterise, on one hand, the airflow that is released from the mouth and, on the other, the initial size distribution of the saliva droplets that are exhaled at the same time.
Both information can be obtained from the currently available data~\cite{yang2007size,xie2009exhaled,gupta2009flow,johnson2011modality,kwon2012study}, measured by performing experiments over diverse human subjects with the goal of providing a standard and representative airflow velocity profile and droplet size distribution.
Looking at such evidence, however, remarkable differences are typically found in terms of the exhaled airflow between male and female subjects~\cite{yang2007size,gupta2009flow,kwon2012study}. E.g., Ref.~\cite{kwon2012study} reported that the initial coughing velocity is about $40\%$ larger for men and observed similar differences for talking, while Ref.~\cite{gupta2009flow} performed a detailed analysis of the time-varying flow rate of cough highlighting the same quantitative discrepancy.
Note that, on the other hand, such characteristic difference is not observed concerning the initial droplet size distribution~\cite{yang2007size,xie2009exhaled,johnson2011modality}.

Motivated by such intriguing issue, here we present a numerical study on the dispersion of droplets released when coughing with emphasis on the characteristics associated with the gender of the exhaling subject.
In our previous work~\cite{rosti2020turbulence,rosti2020urgent} we have considered such kind of expiratory event in light of its importance for the effective transmission of virus-laden droplets, focusing on the accurate description of the physical process governing the transport and evaporation of droplets~\cite{rosti2020turbulence} as well as the influence of the droplet size distribution and environmental conditions on the prediction of the final reach by direct (or short-range) transmission~\cite{rosti2020urgent}.
We now turn our attention into another aspect that can have a relevant role in the problem: the variability with respect to the human subject.

A note should be added with respect to the objective of the present work. Here, our attention is devoted to the droplet exhalation, transport and evaporation by coughing, which is generally referred to as direct (or short-range) transmission, as opposed to indirect (or long-range) transmission associated with aerosol generated especially while breathing or speaking. Recent findings suggest that indirect transmission is the primary driver of transmission in the case of SARS-CoV-2~\cite{morawska2020time,lewis2020coronavirus,asadi2020coronavirus}. Remarkably (and as it will shown in this work), along with a prevailing contribution in terms of direct transmission, to some extent coughing (and sneezing) will also act as long-range transmission~\cite{bourouiba2014violent}.
{Furthermore, it has to be noted that the present analysis is carried out under some simplified conditions (as detailed in the following).}
 
 The remainder of the paper is structured as follows. In Sec.~\ref{sec:formulation}, we introduce the governing equations and how these are solved numerically. Then, in Sec.~\ref{sec:result} we present the main results of our analysis. Finally, in Sec.~\ref{sec:conclusion} we draw some conclusions and perspectives.

\section{Problem formulation} \label{sec:formulation}
\begin{figure}
  \centering
  \includegraphics[width=0.7\textwidth]{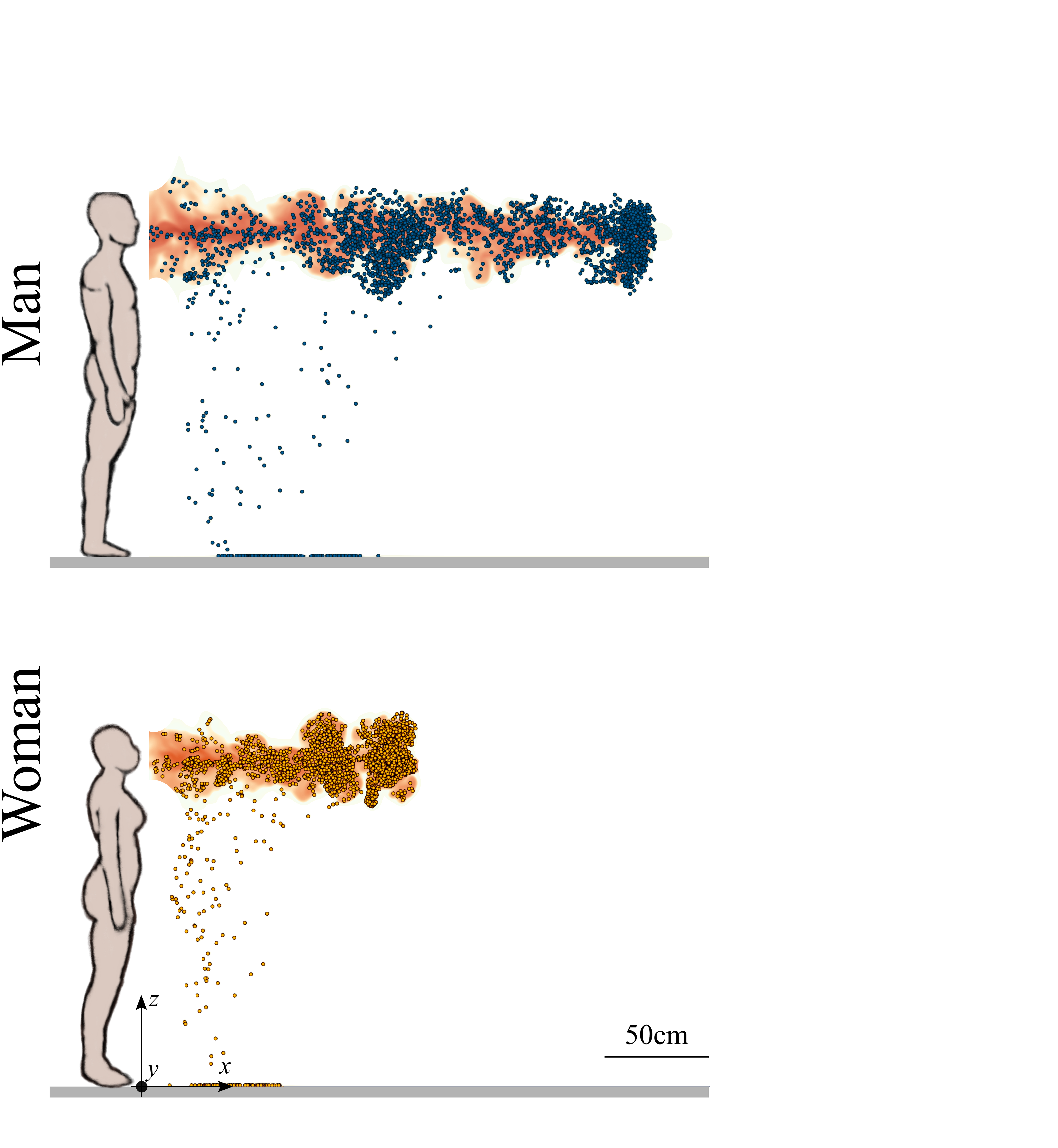}
  \caption{Side view of the instantaneous relative humidity field (color coded) and exhaled droplet positions (blue and orange spheres, not in scale) after $7.6\,\mathrm{s}$ from the starting of cough, for both the male (top) and female (bottom) subject case.} 
  \label{fig:sketch}
\end{figure}

\subsection{Airflow dynamics}
\begin{figure}
  \centering
  \includegraphics[width=0.49\textwidth]{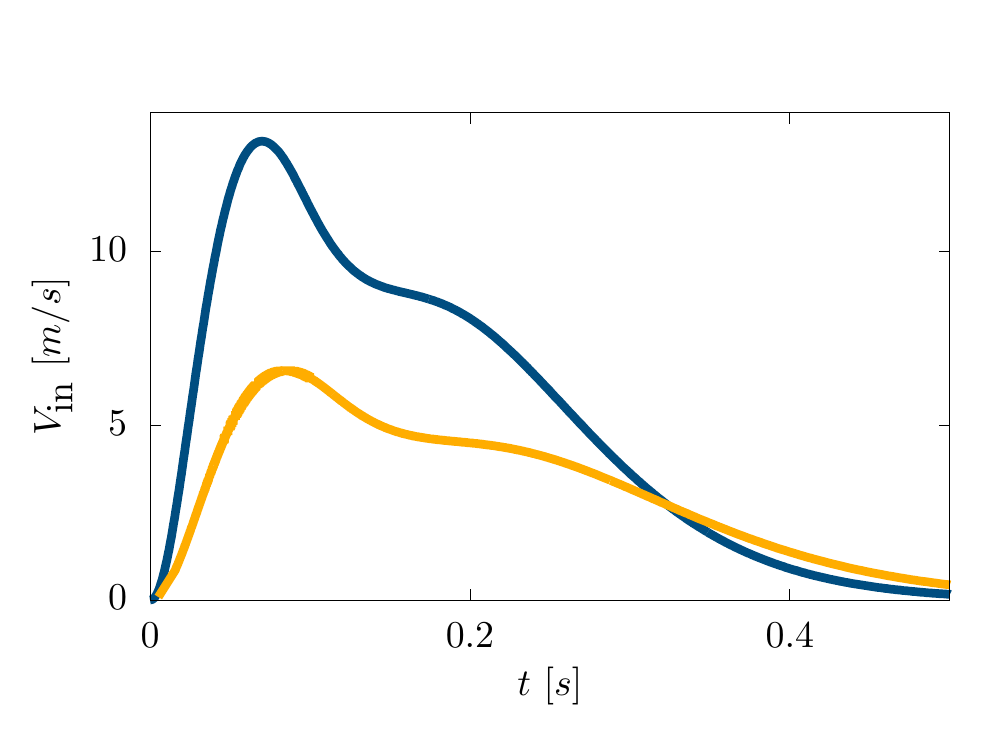}
  \includegraphics[width=0.49\textwidth]{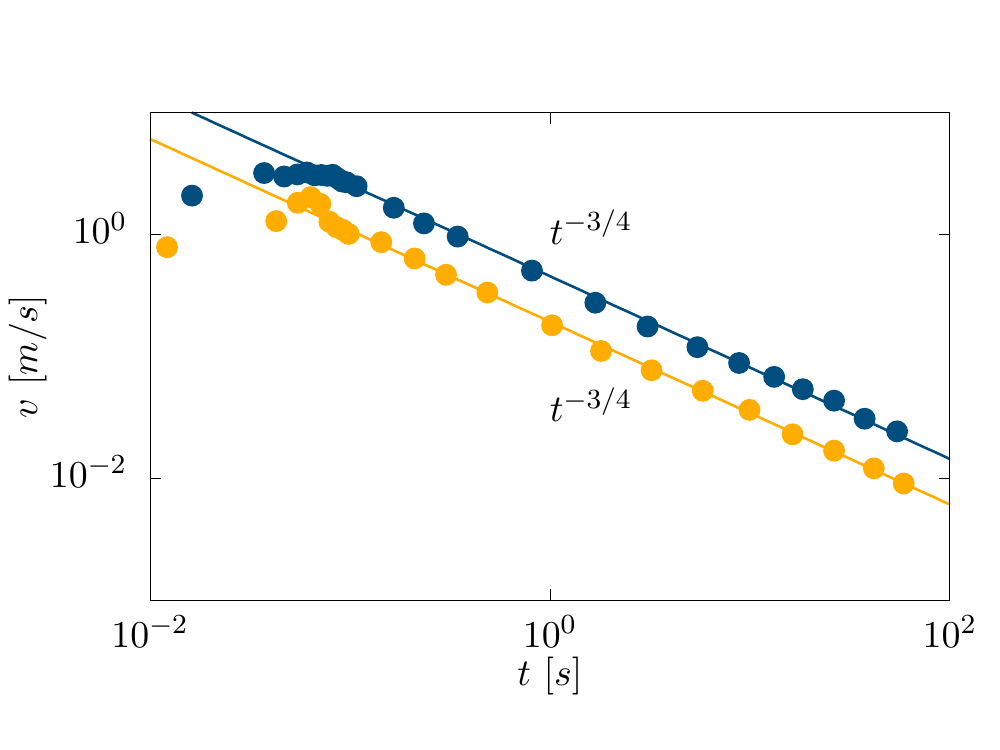}
  \caption{(Left) Time-varying inlet airflow velocity of cough according to Ref.~\cite{gupta2009flow}. The blue and orange lines are used to distinguish the velocity profiles of a male and female subject, respectively. (Right) Time history of the mean streamwise velocity component $v$. The lines show the expected scaling while the symbols are the results of our simulations.}
  \label{fig:vel}
\end{figure}

We consider the airflow generated during a violent expiratory event such as coughing into a quiescent environment~\cite{gupta2009flow}. The flow is governed by the incompressible Navier--Stokes equations
\begin{align}
  \partial_t \bm{u} + \bm{u}\cdot \bm{\partial} \bm{u} &=-\frac{1}{\rho_a}\bm{\partial}p + \nu_a \partial^2 \bm{u}, \\
  \bm{\partial}\cdot \bm{u} &=0,
\end{align}
where $\bm{u}$ and $p$ are the velocity and pressure fields, respectively, $\nu_a=1.8\times 10^{-5}\, \mathrm{m^2/s}$ is the kinematic viscosity and $\rho_a=1.18\, \mathrm{kg/m^3}$ is the density of the air. The airflow emitted from the mouth is more humid than the ambient condition; instead of focusing on the evolution of the relative humidity field $\mathrm{RH}$, we evolve the supersaturation field $s=\mathrm{RH}-1$, which is governed by a simple advection-diffusion equation \cite{celani2005droplet}
\begin{equation}
  \partial_t s + \bm{u}\cdot \bm{\partial} s= D_v \partial^2  s,
\label{eq-supersat}
\end{equation}
where $D_v = 2.5\times 10^{-5}\, \mathrm{m^2/s}$ is the water vapor diffusivity. Eq.~\eqref{eq-supersat} is valid when the saturated vapor pressure is constant, an assumption that holds true as long as the ambient and the exhaled air temperature do not differ substantially.
{Note that in this simplified framework, the supersaturation field behaves passively without any dependence from the fluid temperature.
 In our simulations, we assume the ambient temperature to be $25\,^oC$ and a temperature of the exhaled air of $30\,^oC$~\cite{morawska2009size}.}
  The previous set of equations are solved numerically within a domain of length $L_x=4\,\mathrm{m}$, width $L_y=1.25\,\mathrm{m}$ and height $L_z=2.5\,\mathrm{m}$. The fluid is initially at rest, i.e. $\bm{u}(\bm{x},0)=\bm{0}$, and at the ambient supersaturation $s(\bm{x},0) = s_a=\mathrm{RH}_a-1 = -0.4$ (i.e., considering the environment at a relative humidity of $60\%$). The exhaled air is assumed to be fully saturated \cite{morawska2009size} (i.e.\ $s_\mathrm{mouth}=0$) and is injected through a round opening of area $A_\mathrm{mouth}=4.5\,\mathrm{cm^2}$ mimicking the mouth, at a distance from the ground of $z_\mathrm{mouth}=1.6\,\mathrm{m}$. The injected airflow is prescribed according to the experimental measurements reported by Ref.~\cite{gupta2009flow} and shown in the left panel of Fig.~\ref{fig:vel}. The duration of the exhalation is around $0.5\,\mathrm{s}$ with a peak velocity of $13\,\mathrm{m/s}$ for the male and of $7\,\mathrm{m/s}$ for the female subject. The resulting Reynolds numbers (based on the peak velocity and on the mouth average radius) are about $9\times10^{3}$ and $5\times10^{3}$, respectively. For the other domain boundaries, we prescribe the no-slip condition at the bottom ($z=0$) and left ($x=0$) boundaries and the free-slip condition at the top boundary ($z=L_z$), while applying the Dirichlet condition $s=s_a$. For both the velocity and supersaturation field, we impose a convective outlet boundary condition at the right boundary ($x=L_x$). Finally, periodic boundary conditions are enforced in the spanwise direction (i.e., $y=0$ and $y=L_y$).

\begin{table}[h]
\caption{Physical/chemical parameters representative of expiratory events and adopted in the present investigation.}
\label{tab:param}
\begin{tabular}{l|c|c}
\hline
\hline
Density of soluble aerosol part (NaCl) & $\rho_s$ & $2.2\times 10^3\, \mathrm{kg/m^3}$\\
Density of insoluble aerosol part (mucus) & $\rho_u$ & $1.5\times 10^3\, \mathrm{kg/m^3}$\\
Density of dry nucleus & $\rho_N$ & $1.97\times 10^3\, \mathrm{kg/m^3}$\\
Mass fraction of soluble material (NaCl) w.r.t. the total dry nucleus & $\epsilon_m$ & $0.75$\\
Mass fraction of dry nucleus w.r.t. the total droplet & ${\cal C}$ & $1 \, \%$\\
Specific gas constant of water vapor & $R_v$ & $4.6\times 10^2\, \mathrm{J/(kg\, K)}$\\
Heat conductivity of dry air & $k_a$ & $2.6\times 10^{-2}\, \mathrm{W/K\, m}$\\
Latent heat for evaporation of liquid water & $L_w$ & $2.3\times 10^6\, \mathrm{J/kg}$ \\
Saturation vapor pressure & $e_\mathit{sat}$ & $0.616\, \mathrm{kPa}$\\
Droplet condensational growth rate & $C_R$ & $1.5  \times 10^{-10} \,\mathrm{m^2/s}$\\
Surface tension between moist air and salty water & $\sigma_w$ & $7.6\times 10^{-2}\, \mathrm{J/m^2}$\\
Molar mass of NaCl & $M_s$ & $5.9\times 10^{-2}\, \mathrm{kg/mol}$\\
Molar mass of water & $M_w$ & $1.8\times 10^{-2}\, \mathrm{kg/mol}$\\
Number of ions into which a salt molecule dissociates & $n_s$ & 2\\ 
Osmotic coefficient of salt in solution & $\Phi_s$ & 1.2\\ 
\hline
\hline
\end{tabular}
\end{table}

\subsection{Droplet dynamics}
The exhaled droplets are modelled as an ensemble of $N$ spherical particles of radius $R_i$ dispersed within the airflow. Since the droplet volume fraction for coughing is always smaller than $10^{-5}$ (see e.g.~\cite{wang1993settling,bourouiba2014violent}), we can safely neglect the backreaction to the flow and humidity fields. The dynamics of each droplet is thus governed by the well-known set of equations~\cite{maxey1983equation}
\begin{align}
  \dot{\bm{X}_i}&=\bm{U}_i(t)+\sqrt{2 D_v}\bm{\eta}_i (t), \label{eq:max1} \\
  \dot{\bm{U}_i}&=\frac{\bm{u}(\bm{X}_i(t),t)-\bm{U}_i(t)}{\tau_i} +\bm{g},
\label{eq:max}
\end{align}
for $i=1, \ldots , N$, and where $\bm{X}_i$ and $\bm{U}_i$ are the position and velocity vectors of the $i$-th droplet, respectively, and $\bm{g}$ is the gravitational acceleration. The dynamics is affected by a Brownian white-noise process $\bm{\eta}_i$ and by the Stokes time $\tau_i = 2 (\rho_{D\,i}/\rho_a) R_i^2(t) / 9\nu_a$. 
{For the sake of simplicity, no Reynolds-based correction is used for the viscous drag; it was checked a posteriori that such choice is fully justified during the puff phase, when the Reynolds number experienced by the droplets is always sufficiently small, reaching a value up to $\mathcal{O}(10)$ only in the initial jet phase when the droplets are expelled by the mouth.}
In order to define the density of the i-th droplet $\rho_{D\,i}$, we assume the droplet to be composed by a dry nucleus with density $\rho_N$ surrounded by a salty water layer with density $\rho_w = 9.97\times 10^2\, \mathrm{kg/m^3}$. The dry nucleus is composed by a soluble phase (NaCl) and a insoluble phase (mucus), giving an overall density which can be expressed as 
\begin{equation}
\rho_N  = \frac{\rho_u}{1-\epsilon_m[1-(\rho_u/\rho_s)]},
\end{equation}
where $\epsilon_m$ is the mass fraction of soluble material (NaCl) with respect to the total dry nucleus and $\rho_u$ and $\rho_s$ the density of the insoluble (mucus) and soluble (NaCl) parts. Thus, the density $\rho_{D\,i}$ of the droplet can be computed as
\begin{equation}
\rho_{D\,i}  = \rho_w + (\rho_N-\rho_w)\left (\frac{r_{N\,i}}{R_i(t)}\right )^3,
\end{equation}
where the radius of the (dry) solid part of the droplet when NaCl is totally crystallized is given by
\begin{equation}
r_{N\,i}  = R_i(0)\left (\frac{{\cal C}\;\rho_w}{{\cal C}\;\rho_w +\rho_N(1-{\cal C})}\right )^{1/3},
\end{equation}
being ${\cal C}$ the mass fraction of dry nucleus with respect to the total droplet. Finally, the droplet radius $R_i$ evolves according to a condensation model that has been successfully employed in the analysis of rain formation processes~\cite{pruppacher1997microphysics,celani2005droplet,celani2008equivalent,celani2009droplet}, i.e.
\begin{equation}
  \frac{d}{dt} R^2_i(t) = 2 C_R\left (1+s(\bm{X}_i(t),t)-e^{\frac{A}{R_i(t)}-B\frac{r_{N\,i}^3}{R_i^3(t)-r_{N\,i}^3 } } \right ),
\label{eq-radii}
\end{equation}
where $C_R$ is the droplet condensational growth rate defined as
\begin{equation}
C_R=\left[\frac{\rho_w\, R_v\, (273.15+T)}{e_{sat}\,D_v}+\frac{\rho_w\,L_w^2}{k_a\,R_v\,(273.15+T)^2}-\frac{\rho_w\,L_w}{k_a(273.15+T)}\right ]^{-1},
\label{eq:CR}
\end{equation}
being $e_{sat}$ the saturation vapor pressure and $A$ and $B$ two model parameters
\begin{equation}
e_\mathit{sat}= 6.1078\times 10^2\,e^{(17.27\, T/(T+237.3))} \, \mathrm{Pa},
\label{eq:_esat}
\end{equation}
\begin{equation}
A=\frac{2 \sigma_w}{R_v (T+273.15) \rho_w} \qquad \textrm{and} \qquad B=\frac{n_s \Phi_s \epsilon_v M_w \rho_s}{M_s \rho_w},
\label{eq:A}
\end{equation}
where $\epsilon_v= \epsilon_m(\rho_N/\rho_s)$ is the volume fraction of dry nucleus with respect to the total droplet.
Note that in Eqs.~\eqref{eq:_esat}, \eqref{eq:CR} and \eqref{eq:A} the (ambient) temperature $T$ is expressed in degrees Celsius.
The list of physical and chemical parameters involved in the model is completed by Table~\ref{tab:param} which provides the values selected for the present investigation. 

Concerning the initial droplet size distribution, we assume here the one from Ref.~\cite{duguid1946size}, still considered as a reference on the subject.
Accordingly, we consider initial droplet radii approximately ranging from $1$ to $1000\, \mathrm{\mu m}$ with the $95\%$ falling between 1 and $50\, \mathrm{\mu m}$.
Droplets are set initially at rest and randomly distributed within a sphere of radius $1\,\mathrm{cm}$ located inside the circular pipe from which the exhaled airflow is released. The size distribution is the same for the male and female case, this choice being justified since for the droplet size distribution no significant differences between male and female subjects are reported in the current literature~\cite{yang2007size,xie2009exhaled,johnson2011modality}.

\subsection{Numerical discretisation}
The system of equations is solved numerically using the DNS code named \textit{Fujin}. The equations are discretised in space with the (second-order) central finite-difference method and in time with the (second-order) Adams-Bashfort scheme, except for the droplet dynamic equations~\eqref{eq:max1},~\eqref{eq:max} and~\eqref{eq-radii} which are advanced in time using the explicit Euler scheme with smaller substeps to deal with the numerical constraint required by very small droplets. A fast and efficient FFT-based approach is used to solve the resulting Poisson equation for pressure. The solver has been extensively validated in a variety of problems~\cite{rosti_brandt_2017a,rosti2019flowing,rosti_ge_jain_dodd_brandt_2019,rosti2020increase,olivieri2020dispersed}, see also \texttt{https://groups.oist.jp/cffu/code}. In our simulations, the numerical domain is discretised with uniform grid spacing $\Delta x = 3.5 \, \mathrm{mm}$ in all directions, resulting in a total number of approximately $0.3$ billion grid points. The convergence of the results with respect to the numerical and parametric settings was verified against grid refinement, number of sampled droplets and droplet release time~\cite{rosti2020turbulence,rosti2020urgent}.

\section{Results} \label{sec:result}
Fig.~\ref{fig:sketch} shows a typical side view of our results obtained after $7 \mathrm{s}$ from the respiratory event. The figure clearly shows that the resulting droplet dynamics can be macroscopically classified in two distinct behaviors: \textit{i)} some droplets leave the humid and turbulent air puff released from the mouth and travel vertically within the still ambient fluid, eventually settling on the ground at a relatively short distance from their emission point, showing a predominantly ballistic motion; \textit{ii)} other droplets travel for long distance within the humid air forming a cloud of so-called airborne droplets. Although the same general classification can be done for the male and female case, in the latter the distance traveled by both sets of droplets is substantially reduced. To assess the risk of virus transmission, we define the viral load as the ratio between the initial volume of a subset of droplets (\eg settling or remaining airborne) and the total initial volume of all exhaled droplets.

The exhalation process occurring in violent air expulsions is composed by two different regimes. In the early evolution stage, air is injected into the ambient from the mouth and the resulting flow is a jet, whose dynamics is determined by the conservation of the (time-dependent) momentum flux. In the late evolution stage, the cloud stops to receive momentum from the source and freely evolves in the ambient as a turbulent puff. In the puff stage, the momentum of the cloud is constant; in this case, under the hypothesis of self-similarity and following the results by Kovasznay \etal~\cite{kovasznay1975unsteady} we obtain the following decay law for the mean velocity of the puff:
\begin{equation}
 v \propto t^{-3/4}.
\end{equation}
The right panel in Fig.~\ref{fig:vel} shows the mean velocity resulting from our simulations along with the predicted scaling. Differences can be clearly noticed between the male and female case in the initial jet stage (for which we have different inlet velocity profiles), while both cases agree very well with the theoretical decay law from around half second. The different initial jet phase results in a different coefficient for the scaling law, with the male profile following the law $0.45~t^{-3/4}$ and the female one $0.19~t^{-3/4}$.

\begin{figure}
  \centering
  \includegraphics[width=0.49\textwidth]{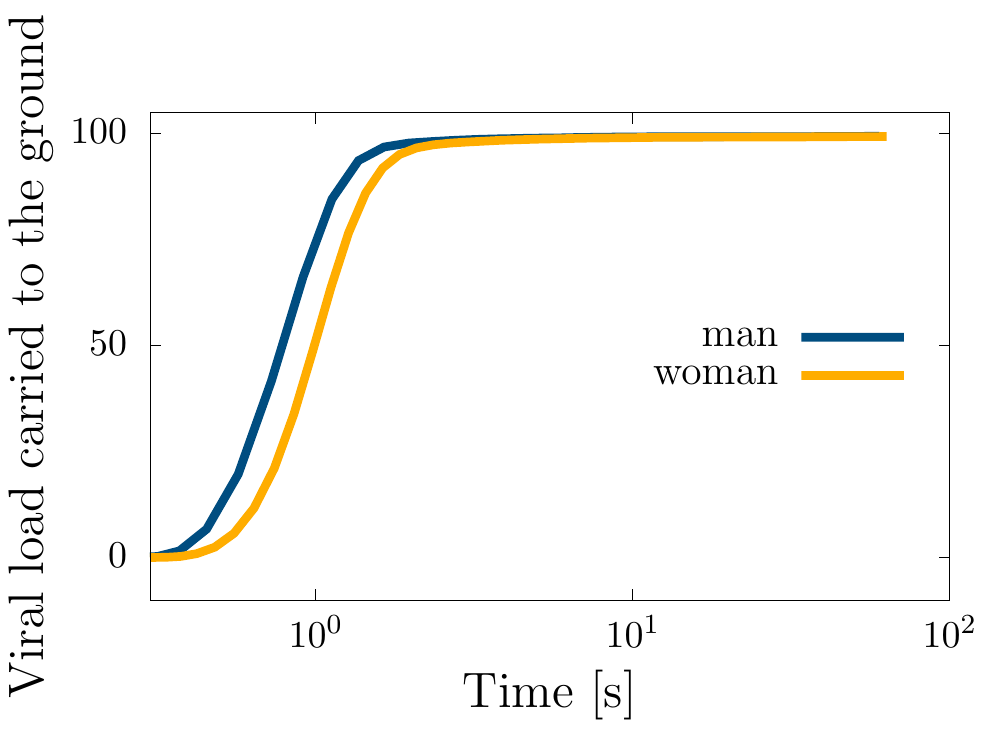}
  \includegraphics[width=0.49\textwidth]{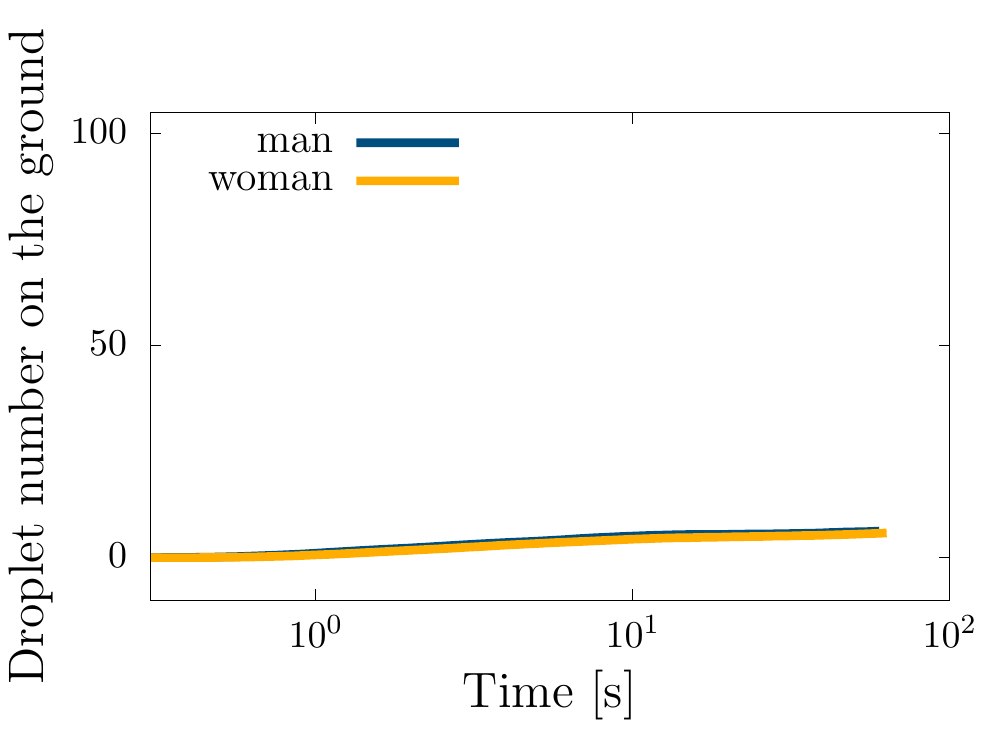}
  \includegraphics[width=0.99\textwidth]{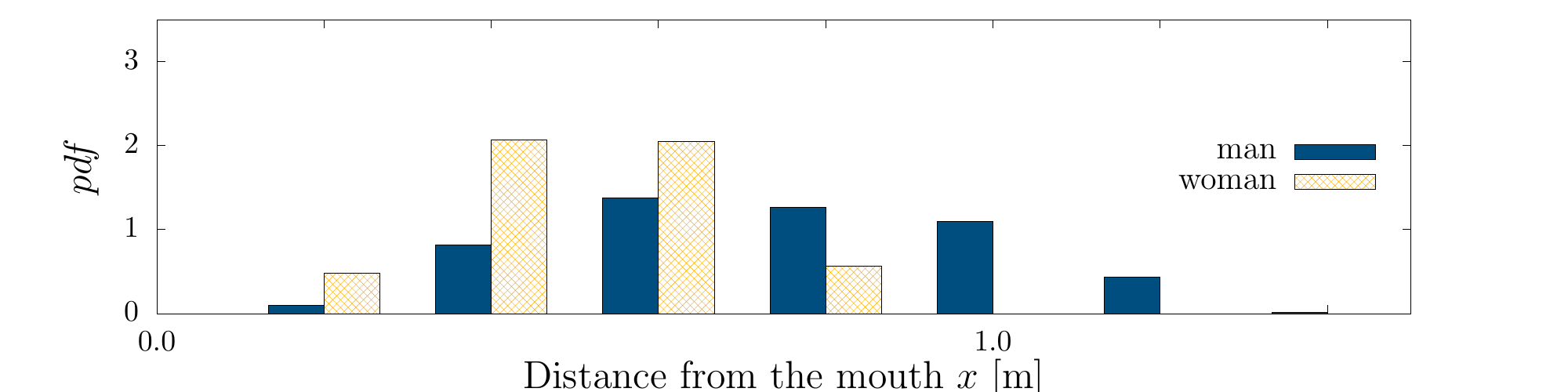}
  \caption{
  Top: Cumulative viral load (left) and normalized droplet number (right) settling to the ground as a function of time (both quantities are expressed as percentage). Bottom: Probability density function of the distance from the mouth when droplets reach the ground. In the figures, the blue and orange colors distinguish the results for male and female, respectively.}
  \label{fig:sedimentation}
\end{figure}

Having characterised the main features of the emitted airflow, we now move to the analysis of the droplet transmission mechanisms, focusing at first on the settling droplets. The left panel in Fig.~\ref{fig:sedimentation} shows as a function of time the viral load of settling droplets. We can observe that in the first few seconds after exhalation, the puff rapidly loses viral load carried by large droplets to the ground. For both the male and female subjects, after around $5\,\mathrm{s}$ approximately $99\%$ of the viral load has reached the ground; this large percentage is however due to a very small number of droplets, around $5\%$, as shown in the right panel of Fig.~\ref{fig:sedimentation}, thus indicating that $95\%$ of droplets remain airborne after $60\,\mathrm{s}$ (note that, at this time, all droplets  are still within the computational domain). The distance reached by the settling droplets is reported in the bottom of Fig.~\ref{fig:sedimentation} where we show the probability density function of the distance travelled by the large droplets when they reach ground. For men, the maximum distance at the ground reached by the large droplets is $1.5\,\mathrm{m}$, reducing to $1\,\mathrm{m}$ for women. It is worth noticing that for these cases the distance is within the social-distancing limits suggested by the World Health Organisation. Note however that, when varying the environmental condition and/or the initial size distribution, such agreement can be lost, showing that the current recommendations may substantially underestimate the safety margin~\cite{rosti2020urgent}.

\begin{figure}
  \centering
  \includegraphics[width=0.49\textwidth]{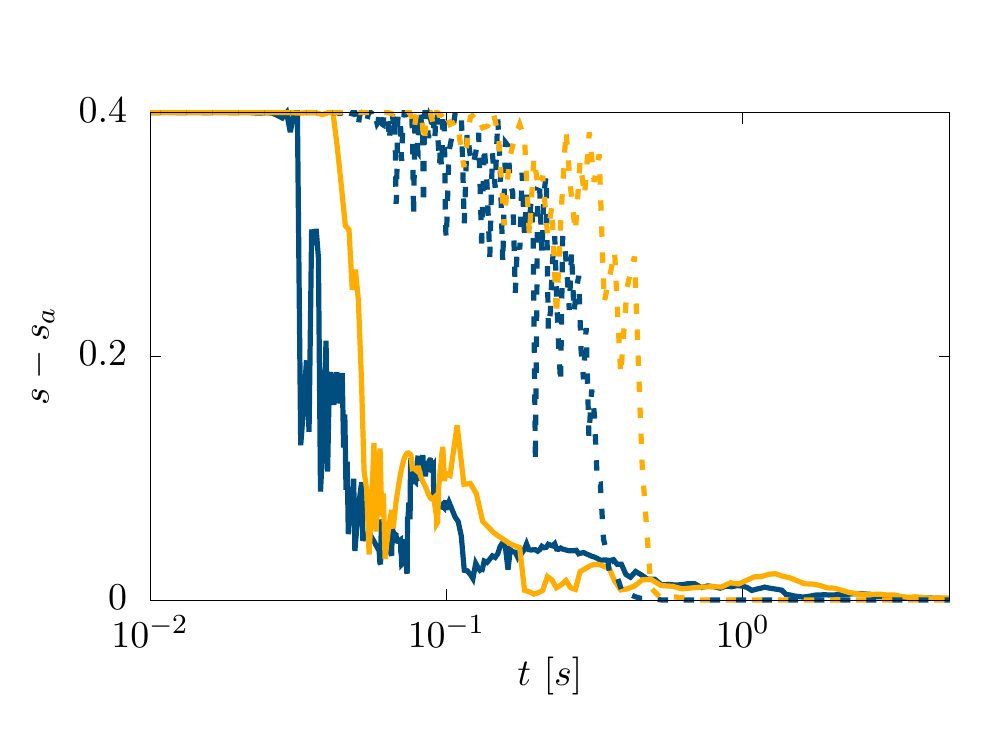}
  \includegraphics[width=0.49\textwidth]{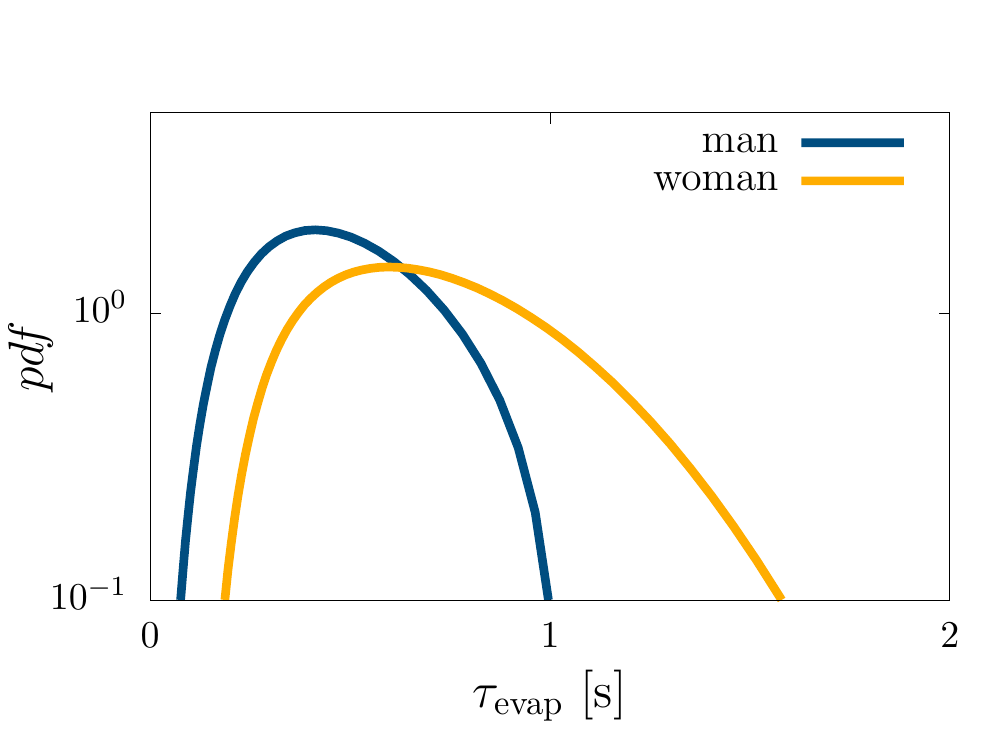}
  \caption{(Left) $s - s_a$ as a function of time experienced by two representative droplets. The solid and dashed lines are used to distinguish a small and large droplet. (Right) Probability density function of the droplet evaporation time $\tau_\mathrm{evap}$, i.e.~the time for the droplet to shrink to its final radius. The observation time is $60\,\mathrm{s}$.}
  \label{fig:evap}
\end{figure}
Let us now focus on the smaller droplets that experience airborne or, more specifically, indirect (or long-range) transmission.
Droplets travel into a non-saturated field and thus evaporate; in Fig.~\ref{fig:evap}-left we report the temporal evolution of the supersaturation field $s$ sampled along the Lagrangian trajectory of a typical small and large droplets. For the large droplet (dashed lines), we observe that the supersaturation field remains similar to the one inside the mouth for long time, indicating an initial slow dynamics due to the droplet inertia, followed by a rapid decay towards the ambient value as soon as the droplet leaves the turbulent puff. On the contrary, the supersaturation field felt by the small droplet (solid lines) shows the so-called ``plateaux-and-cliffs" behaviour, where the scalar field displays dramatic fluctuations occurring in small regions (called cliffs) separating larger areas where the scalar is well mixed (called plateaux). Because of this, small droplets tend to remain long in the large well-mixed regions where they can equilibrate with the local value of supersaturation. The resulting dynamical process is thus made of equilibrium phases alternating with phases of rapid evaporation. Focusing on the influence of the subject gender, the same behaviour can be observed for both men (blue) and women (orange), with the latter showing an overall delay in the dynamics due to the reduced velocity of the exhaled airflow.

The delay in the dynamics for the female subject in turns affects the evaporation process. We thus quantify this feature by measuring the time $\tau_\mathrm{evap}$ needed to reach the final equilibrium size for each airborne droplet, as shown in the right panel of Fig.~\ref{fig:evap}; in particular, we report the probability density function of $\tau_\mathrm{evap}$ for both the male and female case. In general we find broad probability density functions, which is the fingerprint of turbulent fluctuations. However, while for men the mean evaporation time is around $0.4\,\mathrm{s}$, with values ranging between $0.1\,\mathrm{s}$ and $1\,\mathrm{s}$, for women it increases to $0.6\,\mathrm{s}$, with the tail of the distributions even reaching $1.5\,\mathrm{s}$.

\begin{figure}
  \centering
  \includegraphics[width=0.25\textwidth]{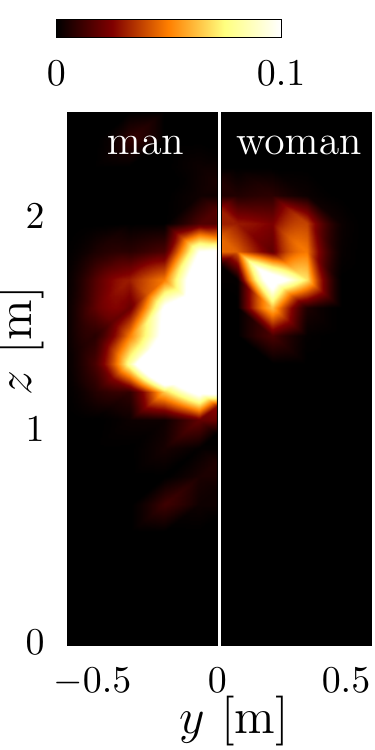} \hspace{1cm}
  \raisebox{0.2\height}{\includegraphics[width=0.49\textwidth]{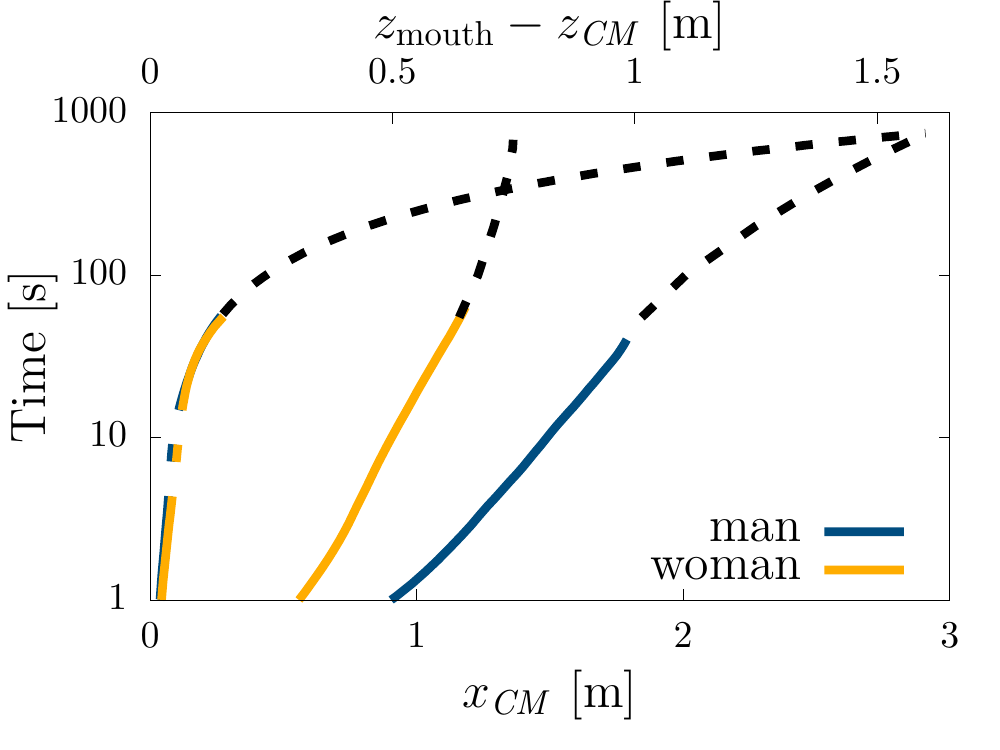}}
  \caption{(Left) Cumulative viral load per unit area reaching a distance of $2\,\mathrm{m}$ from the mouth after $60\,\mathrm{s}$. (Right) Trajectory of the viral load center of mass (computed considering only the airborne droplets): horizontal position $x_\mathit{CM}$ (solid line) and vertical position $z_\mathit{CM}$ (dash-dotted line). The colored lines indicate the results from the simulation while the black dotted ones are extrapolations over longer times.}
  \label{fig:airborne}
\end{figure}
As a further consequence, the observed delay in the evaporation significantly affects the droplet motion. Indeed, we find different predictions for the viral load carried by airborne droplets for the two subjects. In particular, the left panel of Fig.~\ref{fig:airborne} shows the cumulative viral load per unit area reaching the distance of $2\,\mathrm{m}$ from the mouth. In the male-subject simulation, considerable viral load reaches the distance covering an area of around $1\,\mathrm{m}$ in size, while in the female-subject case much less viral load reaches the same distance. To fully quantify the distance travelled by airborne droplets, we track the position of the center of mass of the cloud made exclusively of such subset. For the first $60\,\mathrm{s}$ we can directly compute this quantity from our numerical results while for longer times we extrapolate the trajectory up to the location where the center of mass eventually reaches the ground~\cite{rosti2020urgent}. It clearly appears that, in the absence of external airflow, small airborne droplets can travel several meters: for both men and women, the cloud reaches the ground in about 20 min; due to the different horizontal speed, however, in the female case the  center of mass stops at about $1.5\,\mathrm{m}$ while in the male case it approaches $3\,\mathrm{m}$. 

\section{Conclusions} \label{sec:conclusion}
The present work investigates the influence of the gender of the emitting subject in the dynamics of violent expiratory events in order to better characterise the transport and evaporation process undergone by the exhaled saliva droplets, a topic of paramount relevance for devising improved safety recommendations to face the spreading of airborne infections such as the recent coronavirus outbreak. {To this aim, we focus on the typical cough of both a male and female subject case, and choose a representative condition where the temperature difference between the ambient and exhaled air is such that the supersaturation field can be modelled as a passive scalar. Under such assumptions, a numerical investigation is therefore carried out by means of highly-resolved direct numerical simulations complemented by a Lagrangian model to evolve the droplets released in the flow.}

Among our main findings we observe that, although the physical process is found to be essentially the same from a qualitative viewpoint, substantial quantitative differences occur between men and women in terms of the droplet final reach and evaporation time. The horizontal distance travelled by droplets is found to be generally larger for men, and so is for the cumulative viral load reaching a $2$-meter distance. For women the droplet evaporation time is larger due to a slower dynamics while residing within the turbulent puff. Overall, the results can be associated to the different characteristic airflow, with a typically stronger expulsion for men.

{Our findings further suggest that the current safety guidelines recommending from one- to two-meter-distance are not sufficiently conservative to protect individuals from long-range airborne transmission, either considering typical expiratory events for male or female subjects. The gender of the subject appears as another parameter (along with, e.g., environmental conditions) having an influence on the spreading of virus-containing droplets. However, it should be pointed out that the gender is typically correlated with other physical parameters, such as the weight of the subject. It cannot therefore be excluded at all that the latter has a more direct effect in the observed dynamics. 
Similarly, there are other parameters, such as the age or height of the subject, that can be relevant and should be properly considered to expand the existing knowledge on the topic.
Finally, the present work focuses solely on cough and the extension of our findings to other expiratory events, such as sneeze or talk, is not straightforward due to the peculiar features in the airflow generated by different kinds of exhalation~\cite{kwon2012study}, thus representing an open issue for future investigations. 
}

\section*{Acknowledgments}
The authors acknowledge the computational time provided by HPCI on the Oakbridge-CX cluster under the grant hp200157 of the ``HPCI Urgent Call for Fighting against COVID-19'' and the computer time provided by the Scientific Computing section of Research Support Division at OIST. The research was supported by the Okinawa Institute of Science and Technology Graduate University (OIST) with subsidy funding from the Cabinet Office, Government of Japan. A.M. thanks the financial support from the Compagnia di San Paolo, project MINIERA no. I34I20000380007. Useful discussions with Agnese  Seminara are warmly acknowledged.

\section*{Data Availability Statement}
The data that support the findings of this study are available from the corresponding author upon reasonable request.

\section*{Compliance with Ethical Standards}
Funding: This study was funded by Compagnia di San Paolo (grant number I34I20000380007). 
The authors declare that they have no conflict of interest.

\bibliographystyle{unsrt}

\end{document}